%% file: NoPuzzleMain.tex
\pdfoutput=1 

\documentclass[11pt]{article} 
\usepackage{amsthm, amssymb, amsfonts}
\usepackage[leqno]{amsmath}
\usepackage{newlfont}
\usepackage{graphicx,fleqn}
\usepackage[hang,bf]{caption}
\usepackage{eurosym}        
\usepackage{hyperref}
\usepackage{dcolumn}
\usepackage{rotating}
\usepackage{graphicx,epsf,rotate,rotating}
\RequirePackage{harvard}
\usepackage{harvard}    
\citationstyle{dcu} 
\newcommand{\fn}[1]{\footnote{\hspace{0,5cm}\parbox[t]{14cm}{\setlinespacing{1,00}{#1}}}}
\newcolumntype{d}[1]{D{.}{.}{#1}}
\renewcommand{\thefootnote}{\fnsymbol{footnote}}

\newcommand{\be}{\begin{equation}}
\newcommand{\ee}{\end{equation}}
\newcommand{\bea}{\begin{eqnarray}}
\newcommand{\beas}{\begin{eqnarray*}}
\newcommand{\eea}{\end{eqnarray}}
\newcommand{\eeas}{\end{eqnarray*}}
\newcommand{\caf}{\citeaffixed} 
\newcommand{\can}{\citeasnoun} 
\newcommand{\poc}{\possessivecite}

\newcommand{\multc}{\multicolumn}

\renewcommand{\baselinestretch}{1.5} 

\let\margin\marginpar
\newcommand\myMargin[1]{\margin{\raggedright\scriptsize #1}}
\renewcommand{\marginpar}[1]{\myMargin{\textsf{#1}}}

\newcounter{appeq}
\newcounter{apptab}
\newlength{\defbaselineskip}
\newcommand{\setlinespacing}[1]%
           {\setlength{\baselineskip}{#1 \defbaselineskip}}
\newcommand{\sls}{\setlinespacing}

\begin{document}



\setlinespacing{1.66}%

\thispagestyle{empty}
\begin{table}
\renewcommand{\baselinestretch}{1.0}
\begin{center}
\begin{LARGE}
The puzzle that just isn't\\[9pt]
\end{LARGE}
\renewcommand{\baselinestretch}{1.0}
\begin{tabular}{c}
Christian M\"uller 
\\
\small German University in Cairo\\[-1ex]   
\small Faculty of Management Technology\\[-1ex]            
\small Department of Economics \\[-1ex]
\small 11835 New Cairo City, Egypt \\[-1ex]                       
\small and \\[-1ex]                       
\small Jacobs University Bremen \\[-1ex]                       
\small and \\[-1ex]                       
\small Swiss Ministry of Finance \\[-1ex]                       
\small 8003 Berne, Switzerland \\[-1ex]                       
\small Email: mail@cmueller.ch \\                          
{\sl This version: \today}
\end{tabular}
\end{center}
\end{table}
%
%

\begin{abstract}
In his stimulating article on the reasons for two puzzling observations about the behaviour of interest rates, exchange rates and the rate of inflation, Charles Engel \citeyear{eng:16} puts forward an explanation that rests on the concept of a non-pecuniary liquidity return on assets. Albeit intriguing the analysis struggles to account for a number of facts which are familiar to participants of the foreign exchange and bond markets. Reconciling these facts in conjunction with a careful dissection of the ``puzzle'' to begin with, shows that the forward premium puzzle just does not exist, at least not in its canonical form.

\smallskip
\end{abstract}
{\setlinespacing{1}\noindent {\em Keywords:} foreign exchange rates, forward premium puzzle,
epistemology} 
%


\include{NoPuzzle}


\end{document}

%% file: NoPuzzle.tex


\renewcommand{\arraystretch}{0.5}
\renewcommand{\thefootnote}{\arabic{footnote}}
\renewcommand{\baselinestretch}{2.0}

\begin{flushright}
\begin{minipage}{21pc}
\setlinespacing{1.00}%
 \textit{$\dots$ a scientist must also be absolutely like a child. If he sees a thing, he must say he sees it, whether it was what he thought he was going to see or not. See first, think later, then test. But always see first. Otherwise you will only see what you were expecting.}
\end{minipage}\\[0.5em]
Wonko the Sane 
\end{flushright}

\section{Introduction}
The forward premium puzzle intrigues economists because no matter how hard one tries the data does not live up to a seemingly straightforward, logical theory. Ever since its introduction to the literature \caf{mee:rog:83,fam:84}{e.g.} economists have tirelessly attempted to crack it. The fascinating and stimulating article by \citename{eng:16} in the February 2016 volume of the American Economic Review is another such endeavour.

This note argues that the uncovered interest parity ``puzzle'' is not  a puzzle to begin with, or rather, that the puzzling features are owed to the specific epistemological approach economists have chosen. The argument is twofold. First, the canonical form of puzzle fails to properly account for certain market features although they leave recognisable traces in the standard data sets. Second, and most importantly, the puzzle itself must be considered a methodological construct which is at odds with the very scientific principles that are used in its investigations. 

The arguments are divided into two main parts. Part one deals with the phenomenological aspect of the puzzle and the second with the underlying structure.  

The phenomenological part recalls some stylised facts of the markets under consideration. These markets are the foreign exchange markets and the market for short-term bonds. The according observations will be turned into testable hypotheses shedding light on sometimes disregarded data properties. Checking these properties lends support to rejecting the puzzle. Part two of the argument consecutively adds structural arguments to this note's main claim that the forward premium puzzle is best considered an economists' fiction rather than a real world phenomenon. Finally, conclusions for the future of economic analysis will be drawn.


%
\section{The puzzle that isn't\label{secPzz}}
The starting point for the forward premium puzzle is the well-known no-arbitrage-condition due to the uncovered interest parity (henceforth UIP) condition which can be cast as follows 
$$
i^* - i_t = s_t - s_{t+1}
$$
where $i$ signifies the interest rate on one-period bonds and $s_t$ the log of the price of the foreign currency in period $t$ expressed in home currency units (the US dollar). An asterisk indicates foreign prices. 

The above relationship states that the difference in interest that could be earned by either investing in foreign or domestic bonds must match the change in the log-exchange rate in order to exclude arbitrage opportunities. 

The no-arbitrage condition thus links two otherwise independent markets. It seems so logical and convincing in the light of the efficient market hypothesis that usually only little effort is spent on explaining, let alone substantiating it before turning to the empirical part. However, in order for arbitrage to actually work, some tacit assumptions must hold. For example, it must be maintained that both markets are similar in actual or potential size, or ease of access.  Therefore, we will first look at the actual properties of the markets for gauging the extent to which they are indeed related as the uncovered interest parity condition suggests.

\subsection{The phenomenological approach to puzzle resolution\label{subsecPzzPart1}}
\subsubsection{Stylised facts}
The UIP links two different markets. The no-arbitrage conditions states that they both must influence each other for arbitrage to be ruled out. 

The custom in the literature has it that the short term bond market is captured by the price for money market bonds and the exchange rates are spot rate markets commonly tracked by central banks or other data providers. In the case of \poc{eng:16} analysis the market for one-month money market bonds with prices represented by the mid-point of offer and bid rates of 1-month annual Euro rates are considered. The spot exchange rates in \can[online appendix]{eng:16} are surveyed transaction prices. A first stylised fact therefore is that bond market transaction prices are usually not observed but spot foreign exchange (and foreign exchange swap) transaction prices are.

Second, money market bonds are traded either in the form of swaps or by outright bond purchases. In all cases there is an underlying credit contract between -- usually -- a commercial bank and some other market participant or another commercial bank. Whenever a non-bank is involved at least, the counterparty, the commercial bank will also consider the credibility of the debtor because the contract extends over a non-zero time span. The implication of this observation is that credit contracts are complicated and bear some risk whereas spot foreign exchange contracts do not. It should be expected that this complication puts a brake on the volume of traded contracts.{\fn{Ever since the financial crises, ie after the end of \poc{eng:16} sample, credit swaps also reflect default risks of commercial banks which is a fairly new phenomenon.}} In brief, short-term  bond market transactions are more complicated than spot foreign exchange markets which is the second stylised fact. 

Third, partly owed to the above, partly due to potentially other reasons, markets for money markets bonds are small if not tiny in comparison to spot foreign exchange markets. In fact, it is very difficult to actually measure the size of the market because there are various definitions available. However, irrespective of the definition, spot foreign exchange markets see a turnover that is several multiples that of short-term bond markets (see table \ref{tab1} below). 

The table \ref{tab1} below presents some estimates of the respective sizes based on recent 2013 BIS and Swiss National Bank data. By one such estimate the daily turnover of Swiss Francs against the US dollar and the Euro combined (all foreign exchange instruments) amounted to  USD bill 286 in 2013 while the total of money market instruments by Swiss banks (including local subsidiaries of foreign banks, liabilities plus assets) at the end of 2013 was about USD bill 100.{\fn{Supposing a unit exchange rate for simplicity \cite{snb:15}} In other, words, it would have taken Swiss banks to turn over all their stock of money market liabilities more than twice a day only to reach the volume of the daily CHF/USD and CHF/EUR exchange market. The ratio was smaller in previous years which underlines the liquidity problems money market experienced following the financial crisis. The basic observation of a large gap in market sizes holds nonetheless.  
\begin{table}
  \centering
  \caption{Market sizes}\label{tab1}
  \footnotesize
          \begin{tabular}{lrc}
          Market / instrument & \multc{1}{c}{volume bill USD} & source / remark \\ \hline \hline
          \multc{3}{l}{}\\[-0.5em]
          \multc{3}{l}{Foreign exchange by instrument$^1$ 2013}\\
          outright forward    &       680  / day      & \can[table 1]{bis:13}    \\      
          spot                &     2,046  / day      & \can[table 1]{bis:13}    \\      
          fx swaps            &     2,228  / day      & \can[table 1]{bis:13}    \\      
          currency swaps      &        54  / day      & \can[table 1]{bis:13}    \\      
         options and other$^2$&       337  / day      & \can[table 1]{bis:13}    \\      
          TOTAL               &     5,345  / day      & \can[table 1]{bis:13}    \\      
          \multc{3}{l}{}\\[-0.5em]
          \multc{3}{l}{Foreign exchange turnover by currency pairs$^1$ 2013}\\
          USD / EUR           &      1,289  / day      & \can[table 3]{bis:13}    \\      
          USD / JPY           &        978  / day      & \can[table 3]{bis:13}    \\      
          USD / GBP           &        472  / day      & \can[table 3]{bis:13}    \\      
          USD / CAD           &        200  / day      & \can[table 3]{bis:13}    \\      
          USD / CHF           &        184  / day      & \can[table 3]{bis:13}    \\      
          \multc{3}{l}{}\\[-0.5em]         
          \multc{3}{l}{USD interest rate swaps 6--24 months to maturity$^3$ turnover 2016}\\
          all products        &       1,205 / week    & \can{cft:15}    \\   
          \multc{3}{l}{}\\[-0.5em]
          \multc{3}{l}{Money market instruments all Swiss banks$^4$ 2013}\\
          assets              &         20 by e.o.y   & \can[table 7]{snb:15}    \\      
          liabilities         &         78 by e.o.y   & \can[table 18]{snb:15}    \\      
          TOTAL               &         98 by e.o.y   &          \\ \hline 
          \end{tabular}\\[0.5em]
        \parbox[l]{22.0pc}{\renewcommand{\baselinestretch}{1} \footnotesize
        \sls{1} $^1$Adjusted for cross-border and inter-dealer double counting. $^2$Other includes leveraged products which could not be disentangled into individual plain vanilla products. $^3$Week 28 March - 1 April 2016, includes the following products: fixed float, FRA, Cap/Floor, Debt Option, Exotic, Fixed-Fixed, Inflation, OIS, Swaption, and Basis. Further details: \url{http://www.cftc.gov/MarketReports/SwapsReports/L1INtRatesTransDolVol}. $^4$Stock at end of year.  }
\end{table}

Due to reasons that will be detailed in the second part, being small implies that the expected variance of the price for short-term bonds is small compared to the variance of exchange rates. This expectation is not trivial as one could think that very active markets lead to more efficient price discoveries and hence lower volatility. As a matter of fact, however, foreign exchange markets are much more volatile. For example, the \can{eng:16} data exhibits a variance in the change of the log exchange rate (the right hand side of the UIP condition) that is one hundred to almost three hundred multiples of the variance of the change in the interest rate spread while the correlation is a meagre 3.3 to 15 percent (see table \ref{tab2}).{\fn{To be exact, this comparison is based on the actual regression equation (\ref{eq1}) below. The following claims are not affected by this particular measurement.}} This considerable gap in variances is the fourth stylised fact \caf{wan:08}{see also chart 3 of}. 

Empirically, we thus have to deal with [1] easily observable foreign exchange market prices, [2] very low obstacles to trade, [3] a very large size and [4] high volatility on foreign exchange markets. The short-term bond market by contrast is characterised by not easily observable prices, some obstacles to market access, smaller size and much lower volatility.  

These stylised facts affect the empirical analysis in at least two distinct ways. First, facts [2] and [3] call into question the tacit assumption of equally easy access to the market and the actual possibility to conduct arbitrage. Anyone who would like to invest in short-term bonds instead of foreign exchange would first have to find a suitable counterparty and must undergo a credibility assessment (fact [2]).{\fn{Having to find a perfectly suitable short-term bond does not simplify matters.}} After overcoming these problems, the investor might face too low a volume of bonds available for matching his or her intended investment (fact [3]).

\begin{table}
  \centering
  \caption{FAMA regressands properties: Variances and correlation}\label{tab2}
  \footnotesize
          \begin{tabular}{lr@{.}lr@{.}lcr@{.}l}
          \multc{8}{l}{$\rho_{t+1}   =  \zeta_s + \beta_s\left(  i^*_t -i_t \right) + u_{s,t+1}$} \\
          country & \multicolumn{2}{c}{$Var(\rho_{t+1})$} & \multicolumn{2}{c}{$Var(i^*_t-i_t)$} & \multicolumn{1}{c}{factor}  & \multicolumn{2}{c}{corr. (\%)} \\          \hline \hline
          \multc{8}{l}{}\\[-0.75em]
          CAN           &      3&783   &  0&017 & 229   &    3&74   \\      
          FRA           &     10&248   &  0&065 & 158   &    7&85   \\      
          GER           &     10&596   &  0&050 & 211   &   10&45   \\      
          ITA           &      9&821   &  0&098 & 100   &    3&31   \\      
          JAP           &     11&557   &  0&040 & 286   &   14&95   \\      
          UK            &      9&558   &  0&033 & 290   &   10&50   \\      
          G6            &      6&607   &  0&028 & 238   &    6&84   \\      \hline
          \end{tabular} \\[0.5em]
        \parbox[l]{21.0pc}{\renewcommand{\baselinestretch}{1} \footnotesize
        \sls{1} ``factor'' lists the number with which $Var(i^*_t-i_t)$ has to be multiplied to match $Var(\rho_{t+1})$ where $Var()$ is short for the variance operator. ``corr. (\%)'' reports the correlation between $Var(i^*_t-i_t)$ and $Var(\rho_{t+1})$ in percent. Sources: \can[online appendix]{eng:16}, own calculations.}
\end{table}

Second, due to the difficulties of observing the precise transaction prices on the money market (fact [1]) and the enormous differences in variance (fact [4]), regression analysis must be expected to deliver only imprecise estimates \cite{wan:jon:03}.

Put together, the stylised facts imply that the data of \can{eng:16} will hardly be informative about the presence of the puzzle as claimed on basis of the Fama regressions. Indeed, a careful analysis of the data shows exactly that.

\subsubsection{Fama regressions reconsidered}
The no-arbitrage condition implied by UIP can be re-written as a regression problem. \can{eng:16} uses the excessive return ($\rho_t)$ specification (\ref{eq1}) below
\bea
\rho_{t+1} &    =   & \zeta_s + \beta_s\left(  i^*_t -i_t \right) + u_{s,t+1} \label{eq1} \text{, with }\\
\rho_{t+1} & \equiv & i^*_t+ s_{t+1} - s_t - i_t \nonumber 
\eea
and where $u_{s,t+1}$ represents an error term.

Under UIP we ought to expect $\zeta_s $ and $\beta_s$ to be zero. In contrast to that, \can{wan:jon:03} argue that one should not hold any particular belief about the estimate of $\beta_s$ because it will vary wildly owed to the huge gap in the variances of the right hand side and left hand side of equation (\ref{eq1}). Finally, the forward premium puzzle arises whenever $\beta_s$ is significantly larger than zero. 

\can[pp. 444-5]{eng:16} claims to have found evidence for this puzzle to be present in the data for four out of six currency pairs over the sample June 1979 through October 2009 at the 10 percent level of significance. In view of this evidence he then concludes that the significant deviations from UIP deserve a thorough explanation. 

The following robustness check challenges the finding of sufficient evidence for the forward premium puzzle to prevail. To that aim, the Fama regressions are re-run but with varying sample definitions. Bearing in mind that a puzzle should be found in most circumstances, that is independent of time and space to qualify for a dominant regularity, if not a law that demands a general explanation or resolution, a robustness check in this manner is very desirable. Furthermore, \poc{eng:16} sample excludes the post-crisis years which implicitly suggests that the puzzle probably has not survived the most recent market developments.

We apply two recursive estimation procedures plus a rolling window approach. In all three cases a maximum of five years of data is shed. In the forward recursion we fix the start point of the sample and delete one observation after the other starting with the latest data point. The backward recursion sees the end point fixed and at each step of the recursive estimation a data point at the beginning of the sample is omitted. Finally, the rolling window starts with five years of data deleted from the start of the sample. With each additional estimation, the first sample observation is shifted backwards while deleting one observation at the beginning. We thus keep the number of observations constant in all rolling windows regression while the sample sizes range between the full sample size and the full size less 60 in the recursions. 
\begin{figure}
\begin{center}
\rotatebox{0}{\resizebox{13.835cm}{15.657cm}{\includegraphics[bb=0 0 378 387,scale=1,keepaspectratio=true]{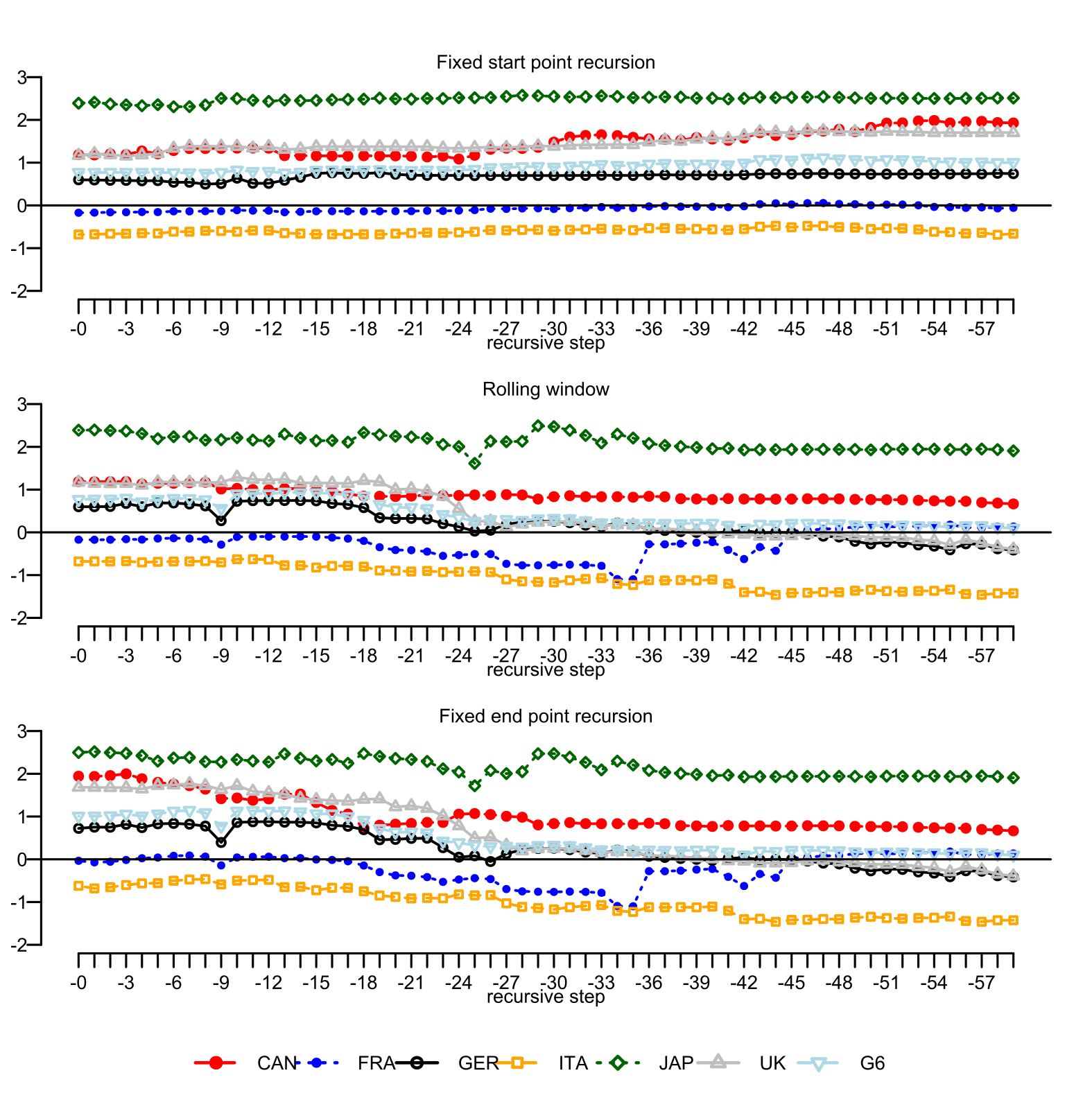}}}\\[0,5ex]
\parbox[l]{30pc}{\footnotesize \sls{1} The figure displays estimated lower bounds for the 90-percent confidence intervals for $\beta_2$ in the regression $\rho_{t+1}   =  \zeta_s + \beta_s\left(  i^*_t -i_t \right) + u_{s,t+1}$. The various points correspond to different sample definitions. Any crossing of the zero line is evidence for non-robustness of the coefficient estimates.}%
\caption{\normalsize Recursive Fama estimations \label{fig1}}%
\end{center}
\end{figure}

The results of the recursive estimations are summarized in table \ref{tab4} and in figure \ref{fig1} below. Starting with the figure, the top panel which displays the lower bounds of the 90-percent confidence intervals for $\beta_s$  obtained with a fixed starting point it seems as if the claim of a generally prevailing puzzle indeed holds true. This can be inferred from the fact that throughout all recursions the lower bounds of the $\beta_s$ coefficients for all countries but one remain either above (Japan, Germany, Canada, United Kingdom) or below (Italy) the zero watershed. The case of France is an exception because it exhibits a brief spell of above zero indicating that the puzzle temporarily prevailed there too. The only slightly irritating feature seems to be the downward trend of the lower bounds. All in all, only little evidence (France) arises against the robustness of \poc{eng:16} Fama regressions.
\begin{table}
  \centering
  \caption{Recursive FAMA regressions: Puzzle on and off by sample definition\label{tab4}}
  \footnotesize
          \begin{tabular}{lr@{.}lr@{.}l}
          \multc{5}{l}{$\rho_{t+1}   =  \zeta_s + \beta_s\left(  i^*_t -i_t \right) + u_{s,t+1}$} \\
           \multicolumn{5}{l}{90\% lower bound  of $\hat{\beta}_{s}$ by sample}  \\
                        &            \multc{2}{l}{1979:6} & \multc{2}{l}{1984:6}  \\    
                country &            \multc{2}{l}{2004:10} & \multc{2}{l}{2009:10}  \\    \hline \hline
          \multc{5}{l}{}\\[-0.75em]
          CAN           &             1&929 &  0&666    \\      
          FRA           &            -0&056 &  0&137    \\      
          GER           &             0&743 & -0&413    \\      
          ITA           &            -0&663 & -1&425    \\      
          JAP           &             2&511 &  1&906    \\      
          UK            &             1&703 & -0&392    \\      
          G6            &             1&004 &  0&096    \\    \hline  
          \multc{5}{l}{}\\[-0.75em]
          \multc{5}{l}{Evidence summary}\\
          \multc{5}{l}{Head count of forward premium puzzle evidence}\\
          supporting    &            \multc{2}{l}{$\,\,$4} &  \multc{2}{l}{$\,\,$3}  \\
          contradicting &            \multc{2}{l}{$\,\,$2} &  \multc{2}{l}{$\,\,$3}  \\[0.25em]
          \multc{5}{l}{Weighted evidence$^1$}\\
          supporting    &             0&72      &   0&44           \\  
          contradicting &             0&28      &   0&56           \\  \hline
          \end{tabular} \\[0.5em]
        \parbox[l]{18.0pc}{\renewcommand{\baselinestretch}{1} \footnotesize
        \sls{1} The head count lists the number of countries for which the specific sample lends support to the forward premium puzzle. The weighted evidence combines the countries and their respective weights as defined in \can[online appendix 1.4.1]{eng:16}. Own calculations based on \poc{eng:16} R code \cite[online appendix]{eng:16}.}
\end{table}

If we move to the centre panel, however, the story changes dramatically. This panel reports the result of the rolling window which implies that the left-most (right-most) data points match the right most (left-most) of its upper (bottom) neighbour. Depending on the specific sample definition, either three countries (France, Canada, Japan) or four countries (United Kingdom, Germany, Japan, Canada) meet the puzzle criteria. What's worse, three countries swap their positions between being affected by the puzzle and being not subject to it (France, Germany, United Kingdom). The switch of Germany and UK is the more significant for the argument as Germany represents the most important foreign exchange market by \poc{eng:16} weights.{\fn{\can{eng:16} assigns Germany and UK a combined weight of 43 percent, with 29 percent for Germany alone.}} Finally, two countries' lower bounds cross the zero line multiple times with Germany three times and France even on no less than ten occasions. 

Given the relative stability seen in the upper panel which relates to the fixed start point recursion and the volatile results for the rolling window, one might expect that the most influential section of the sample must be located in the distant past. The bottom panel supports this conjecture. It shows the recursive estimations with the sample end fixed. The left-most observations consequently resemble the results as of \can[table 1]{eng:16} while the right-most data points correspond to the sample 1984:6 through 2009:10. This shorter sample omits the early Volcker years, for example. Based on this shorter sample, again three countries switch sides (two from puzzle to non-puzzle and France in the opposite direction) leading to a three-three draw by head count. However, the two heavy-weights Germany and United Kingdom do not lend support to the presence of a puzzle neither does a weighted count (see table \ref{tab1}).

Finally, considering the more standard level of significance of $5$ percent instead of $10$ percent further raises the doubts because the conclusion about the presence of the puzzle even more strongly depends on the specific definition of the sample. Canada and Japan are the only two countries for which the presence of the puzzle can be confirmed robustly. Italy, by contrast, is the country for which no evidence can be found and the G6's finding are not robust when looking at the 5 percent significance level.{\fn{\can{eng:16} focusses on the G6 (weighted) average. Focussing on that aggregate might be technically convenient, but from the viewpoint of actual arbitrage it certainly is the most problematic example.}}

To sum up the robustness check, it must be contended that there is no conclusive evidence about the presence of the forward premium puzzle. The findings of \can[p.444]{eng:16} are most probably  driven by the most distant parts of the sample. Looking at slightly more recent evidence the only admissible conclusions appears to be that in some cases (countries) the puzzle might have been present for some periods, while not so for other countries and not in other periods. The fact that in many cases the lower bounds cross the zero line multiple times also safeguards against possible sample size effects. When shedding observations one can expect that confidence bands grow wider. The recursions do show, however, that there is no monotonous trend in the evolution of the confidence bounds.{\fn{The additional recursions presented in the appendix convey the same message.}}

For just two countries robust evidence in favour and for one country against the puzzle exists. The former group does not include the most important country by weight but Canada and Japan which are clearly distinct from the rest of the countries by extreme proximity (Canada) and a very particular economic and monetary situation (Japan). To put it differently, it is impossible to accept the notion of the forward premium puzzle to be anything close to a general economic law. The question hence arises what investor would have dared to bet on ``excessive'' returns based on the presented evidence. The doubts are the more significant as the most relevant observations in support for the arbitrage opportunities are from the late 1970ies, early 1980ies but not from data close to the time of analysis which is from 2009.

An extended recursive analysis that also cover the central empirical findings is presented in the appendix (see pp. \pageref{secAppRecs}f). It shows that virtually all relevant empirical findings can easily be reversed with an according re-definition of the sample.

Therefore, the robustness check confirms the earlier statement about the implications of the stylised facts for the possibility of obtaining conclusive evidence about the forward premium puzzle. As it turns out, it is impossible to draw definite conclusions. In other words, there might be a puzzle or there might not be, we just cannot tell.

\subsection{The structural approach to puzzle resolution\label{subsecPzzPart2}}
\subsubsection{The fiction of a forward premium puzzle\label{subsubsecfict}}
Given the shaky foundations on which the empirical proof of the forward premium puzzle rests, there appears no urgent need to actually elaborate any further on the reasons for the ``puzzle'' as its very existence must be doubted in the first place.{\fn{Any such research could, however, shed light on the special cases of Canada and Japan.}} In the following and in line with the literature, this note maintains that it is impossible to empirically decide about the existence of a puzzle with sufficient certainty and presents some structural reasoning for why this might be the case. 

The explanation rests on two pillars. The first concerns the actual generation of market prices and the second the discussion about the laws which are supposed to govern these prices.

As an illustration of these two pillars consider \can{jor:tay:12} who devise a profitable arbitrage strategy based on carry trades. To that aim, \citename{jor:tay:12} take a stance on what determines foreign exchange rates, among other things. Their stance must be novel as compared to previous stances because their work would not have been fit for publication (or their strategy would have been taken advantage of which would ultimately render it useless due to copying). Suppose now that they would actually implement their ideas. With each market transaction  based on their strategy a new price observation will arise. This particular observation will hence be ruled by their novel idea (and by another strategy that devises the same price). Hence, the according price impulse follows the rules of \poc{jor:tay:12} model. 

The literature on exchange rates is vast, however. Add to this that many strategies never see the light of the day as they are well-kept secrets of professional traders. Only a very small portion of these ideas will be similar to one another because they could otherwise not be used by traders{\fn{By the law of efficient markets any publicly known profitable trading strategy must vanish.}} (nor be accepted for publication for that matter). Therefore, the market is populated by agents who all hold more or less slightly but genuinely  differing views about what prices to trade on. For example, a statement by a government official, a terrorist act and so on will be interpreted by traders but each trader would give it a slightly different interpretation with only the initial direction of change to be agreed on in the best of cases.

Coming back to the data generation we must content that every \textit{actual} observation is driven by the views of ultimately idosynchratic agents quite like in that instant when \citename{jor:tay:12} enter the market. The market price can thus be compared to a ball that receives a kick and each kick is different from the previous one making its movement unpredictable. On top of this, the more hits the ball takes the more wildly will it move around.{\fn{Note the similarity and difference to the random walk view. Random walk shocks are also unpredictable but assumed to follow a statistical law which does not apply here.}} 

Acting along the lines of ones' model thus reveals the \textit{formative} nature of foreign exchange (and many asset) prices. This feature sets those markets apart from problems in science where the researchers' ideas do not have an impact on the event under study such as the orbit of planets.

According to stylised fact [3] foreign exchange markets are huge. To the extend that size also approximates trader heterogeneity, stylised fact [4] which emphasizes the large volatility of foreign exchange markets, is a direct consequence of [3]. Therefore, [3] and [4] are nothing but two sides of the same coin.

\subsubsection{The epistemology of the forward premium puzzle\label{subsubsecEpist}}
In order to fully comprehend the nature of the forward premium puzzle another pillar enters the argument. This pillar concerns the pattern of economists' discussion which is best characterised by the search for a unifying theory for, for example, the determination of asset prices. This intention can be justified by the analogy to science where, for instance, all the details of the laws that govern the orbit of the earth have been discovered with the same \textit{positivist} approach: first assume the existence of a universal law and then scrutinize data, reject faulty laws, propose a new law and so on until the true law is established beyond doubt.{\fn{Arguably, many scientist would certainly object to this description as they take pride in proving theories in a strictly mathematical, axiomatic way. However, experimenting and predictions essentially are positivist approaches.}} In the context of social sciences and under the assumption of a universal law, \can{mut:61} has demonstrated that rational reasoning of agents will eventually lead to its discovery. Needless to say, econometric analysis also ultimately rests on this chain of argument.

The first pillar of the structural explanation of the puzzle has established, however, that it is close to impossible that a universal law actually drives markets and generates observable data. Therefore, the custom of economic analysis is responsible for finding the forward premium ``puzzle'' which otherwise does not exist. Many puzzles, in other words, only exist on the drawing boards of a certain, albeit dominant, approach to economic analysis and it is highly questionable that they have any bearing for the real economy. 

It probably has, however, considerable bearing for economics. This is because as a consequence of the above the forward premium puzzle consists of two inseparable components. They are the existence of an universal law on the one hand and the prevalence of rational behaviour on the other. If either of both was absent then the puzzle would falter. Unfortunately, any empirical test that shows a contradiction between theory and data such as Fama regressions overwhelmingly do{\fn{The theory says that $\beta_s$ is exactly zero and $\zeta_s$ too.}} will offer no hint as to which of the two building blocks is wrong.

Though it is beyond the scope of this note, a brief look at economists' handling of the forward premium puzzle is warranted. In practice, economists have mainly tried to reconcile the forward premium puzzle by either of the following: Amending the underlying theory, establishing or justifying irrational behaviour, using superior data, allowing for heterogeneous agents (models) of sort. 

All of these directions of research except for the last one firmly rest on the assumption of a single, valid underlying model. But even in the case of heterogeneity this heterogeneity is usually bounded which makes it essentially equivalent to the single model case with a sufficiently large sample. The true challenge, however, remains to capture the situation in which the number of genuine models grows largely proportionate to the number of agents because only in this situation proper account is taken of the factually formative nature of exchange rate modelling. 

Maybe ironically, by their very, tireless hunt for puzzle resolutions, by jumping back and forth from amending the theory to fixing rationality to allowing for heterogeneity and many more, economists have long left behind the positivist approach to knowledge generation they have once embarked on. A positivist approach would require to reject a theory once it does not comply with the data quite like Newton's law of gravity ultimately had to be rejected. The incomprehensively large literature on the forward premium puzzle would qualify for a rejection as well. Against all odds, however, economists hold onto it. 

This reluctance to hypothesis rejection leads to the interesting situation that quite many versions of the forward premium puzzle / UIP co-exists. They are distinct from one another by their degrees to which deviations from rational behaviour are allowed, what data is used, and many variations of the core formulae and additional variables like in \can[sec. C.]{eng:16}, for example. When many theories and explanations of the same facts are held valid then the according scientific approach is called \textit{constructivism} as the truth is the result of the scientist's constructs rather than an objectively determinable fact. In that sense, failing to reject UIP outright and piling up UIP variation after variation exactly mirrors the formative nature of asset price generation.

\subsection{Some implications for economic analyses\label{subsecImplis}}
The key distinction between positivism and constructivism rests with the properties of the assumed truth to be discovered. In the former case, it is unique, objective and hence independent of the researcher's judgement whereas in the other it depends on human action because it is established by it. 

If the positivist approach was more appropriate than the constructivist then the following must also be true. As there presumably exists only one, objective truth, the claim of having found it should be corroborated with appropriate experiments. 

For example, claims of feasible excess returns, like the forward premium puzzle, must be proven by according actual evidence. This would entail the documentation of real, excessively profitable investments. It is not sufficient to claim the existence of abnormal profits without operating on them because it would lack the proof of these profits to exist independently of ones' own judgements. In other words, it must be ruled out that the researcher's action has formative character for the market under consideration. 

There are also a number of approaches which are doubtful in the context of a potentially constructivist nature of the problem. For example, tests of rationality are meaningless whenever the researcher is able to determine the objectively correct result of a set problem. If test persons fail to discover the truth it will have no bearings on a real world that is defined by multiple, potentially infinite such truths and even more so if the truth was formative in nature. 

If economists would openly embrace the possibility of a constructivist approach that would push wide open the door to answering a myriad of research questions. For example, what leads individuals to follow individual strategies instead of lining up for a common blueprint? Why do rationality tests (sometimes) fail even in the presence of objective truths?  What are the consequences for macro models? Is the formative nature endogenous or exogenous? Can the constructivist approach be reconciled with the concept of efficient (financial) markets?

In other words, instead of spending another thirty-odd years on waging the same wars on an empirically fragile no-arbitrage condition economists might instead take full advantage of the opportunities the new approach affords them in order to shed a new, bright light on understanding actual market dynamics.

\section{Conclusions}
The forward premium puzzle states that there are excess returns on financial markets which would be at odds with the standard efficient market hypothesis. This comment shows that the claim of having found statistical evidence for the forward premium puzzle to prevail rests on very shaky grounds. In fact, the results of several recursive estimations show that the findings by \can{eng:16} to a large extent depend on the choice of the sample with several countries switching their roles as evidence in favour and against the puzzle. This observation renders the claim of excessive returns rather coincidental than systematic though the well-documented finding of implausible parameter values for the most important point estimate remains.

As a consequence of the inconclusive attempt to establish the forward premium puzzle this comment suggests to consider a change in the epistemological approach to the UIP in particular and market dynamics at large. The proposed constructivist view promises a great potential to significantly augment our understanding of actual market dynamics.


\appendix

\section{Additional recursive analyses\label{secAppRecs}}
In the absence of an objective stochastic process the constructivist approach to economics seems appropriate. Econometric methods based on the positivist view are, by contrast, bound to fail to produce reliable, robust evidence for whatever theory is put forth. Therefore, the FAMA regressions presented in \poc{eng:16} were expected to deliver only fragile, inconclusive evidence. Figure \ref{fig1} above shows the results of standard robustness checks by systematically varying the sample size in the FAMA regressions.

The following figure demonstrates that the lack of robustness must also be conceded for the other key regressions (equations (7), (8) and (9)) reported in tables 3 through 5 respectively in \can{eng:16}. To that aim the sample start is varied between June 1979 and June 1984. Obviously, a conclusion drawn from a sample that covers 1979:6 through 2009:10 should be identical to the conclusion based on the sample 1984:6--2009:10 or any sample in between. Minor deviations of the results should be allowed for in one out of ten cases or so, depending on the actual level of significance.

The below figure shows, that these deviations are way too frequent, to be acceptable by any meaningful margin of error. The panels depict the lower bounds for the confidence intervals of the key coefficient estimates based on the interval bootstrap procedures. The confidence intervals cover 90 percent of the possible true parameter values according to the respective estimations.{\fn{The technical details are provided in \can{eng:16}.}} All coefficients must to be positive with the exception of equation (9) for supporting the claims of the paper. Graphically, this means that their lower bounds are supposed to be above the zero line. In equation (9) the key coefficient should be negative which is why the lower bounds are multiplied by minus one in order to simplify the interpretation of the corresponding (bottom) panel. 

\begin{figure}
\begin{center}
\rotatebox{0}{\resizebox{13.835cm}{15.657cm}{\includegraphics[bb=0 0 504 504,scale=1,keepaspectratio=true]{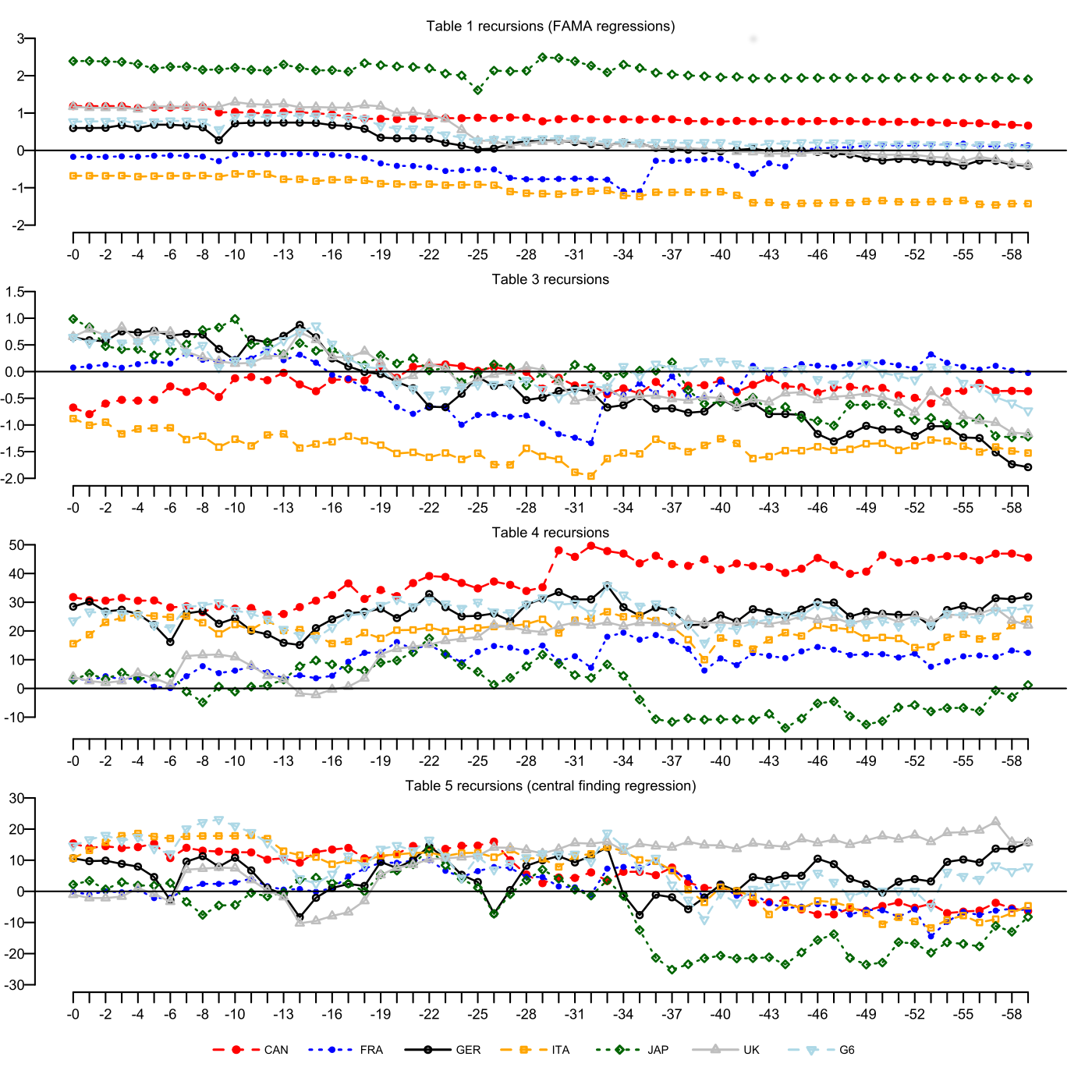}}}\\[0,5ex]
\parbox[l]{30pc}{\footnotesize \sls{1} The figure displays estimated lower bounds for the 90-percent confidence intervals for the four key coefficients in \poc{eng:16} regressions (4), (7), (8), (9) (from top). The various points correspond to different sample definitions. Any crossing of the zero line is evidence for non-robustness of the coefficient estimates. The lower bound for the key coefficient in (9) is multiplied by $-1$. Equations (7), (8), (9): bootstrapped percentile intervals.}%
\caption{\normalsize Some more recursive analyses \label{figA1}}%
\end{center}
\end{figure}

Apparently, with the exception of equation (8) which corresponds to the results in \poc{eng:16} table 4, all regression results are very much dependent on the specific sample definition. To see that, consider first the left-most points in all panels. They represent the results obtained in \citename{eng:16}'s article. Obviously, in the majority of cases the lower bounds are above zero and hence support \citename{eng:16}'s line of arguments. As we drop one observation after the other, the evidence deteriorates, however. Arguably, the second panel from the top shows the worst results. This panel checks the robustness of the FAMA regressions in real terms. With the full sample range (364 observations) the forward premium puzzle in real terms finds support with the exception of Canada and Italy (10 percent level of significance, interval bootstrap). Dropping 60 observations from the beginning of the sample (leaving 304 observations) makes all confidence intervals include the zero. Therefore, based on the sample 1984:6--2009:10 the forward premium puzzle in real terms disappears entirely. 

The ``central empirical finding'' (p.450) of \can{eng:16} is scrutinised in the bottom panel. Again, changing the sample definition topples the evidence. While the 1979 -- 2009 sample support the main finding  at large (all cases except France and UK), the sample 1984 -- 2009 clearly contradicts it (with the exception of -- again -- UK and Germany). In between, some countries' lower bounds drop below zero and bounce back and vice versa quite often. These multiple crossings of the zero line also safeguard against possible sample size effects. Dropping some observations should, in principle, widen the confidence bands but in the present situation that does not lead to a systematic loss of significance.

In summary, the empirical evidence presented in \can{eng:16} must be considered very fragile. Choosing the paper's particular sample definition certainly permit the conclusions drawn by the author. In the light of more recent samples the whole argument falls apart, however.

\bibliography{chris}
